# Accurate Critical Exponents from the Optimal Truncation of the $\varepsilon$-Expansion within the $O(N)-$symmetric Field Theory for large $N$

Abouzeid M. Shalaby[*]

*Physics and Materials Sciences Department, College of Arts and Sciences,*

*Qatar University, Al Tarfa, Doha 2713, Qatar*

The perturbation series for the renormalization group functions of the $O(N)-$symmetric $\phi^4$ field theory are divergent but asymptotic. They are usually followed by Resummation calculations to extract reliable results. Although the same features exist for QED series, their partial sums can return accurate results because the perturbation parameter is small. In this work, however, we show that, for $N \geq 4$, the partial sum ( according to optimal truncation) of the series for the exponents $\nu$ and $\eta$ gives results that are very competitive to the recent Monte Carlo and Conformal field calculations. The series can be written in terms of the perturbation parameter $\alpha = \sigma\varepsilon = \frac{3}{N+8}\varepsilon$ which is smaller for larger $N$ and thus as $N$ increases one expects accurate perturbative results like the QED case. Such optimal truncation, however, doesn't work for the series of the critical exponent $\omega$ ( for intermediate values of $N$) as the truncated series includes only the first order. Nevertheless, for $N \geq 4$, the large-order parameter $\sigma = \frac{3}{N+8}$ is getting smaller which rationalizes for the use of the Padé approximation. For that aim, we first obtain the seven-loop $\varepsilon$-series from the recent available corresponding $g$-expansion. The seven-loop Padé approximation gives accurate results for the three exponents. Besides, for all the seven orders in the series, the large-$N$ limit leads to the exact result predicted by the non-perturbative $1/N$-Expansion.

---

[*] amshalab@qu.edu.qa

# I. INTRODUCTION

Second order phase transitions in many systems can be described by an effective O(N)-symmetric field theory [1–4]. Near critical points, quantities like susceptibility and correlation length diverge and perturbative Renormalization group (RG) are believed to give false results. This is why the RG perturbation series is always followed by a resummation technique [1, 2, 5–7]. Other non-perturbative techniques like the $1/N-$Expansion [4, 8–12] have been used successfully to investigate the theory specially for large $N$, where $N$ is the number of fields. This technique in fact gives exact results for the large-N limit. Recently, Monte Carlo calculations have been carried out for the model for $N \geq 4$ [3]. Also, the model under consideration has been subjected to conformal field studies which can be found in the review article in Ref. [13] and the references therein.

The large-N expansion is convergent within its radius of convergence where it can be summed to give reliable results especially for large $N$ and one can find in literature a plenty of resummation of the first few orders of the $1/N$ expansion [14–17]. On the other hand, the RG $\varepsilon$-expansion ($\varepsilon = 4 - d$) is divergent with zero-radius of convergence [1, 2, 6, 7, 18, 19]. This is why such series (without resummation) are always overlooked in literature when one is seeking accurate results for the critical exponents (for instance). However, the series is asymptotic in the sense that there exists a minimum order after which no more orders can add an improvement to the results. Because of divergence, it is always believed that the large-N expansion is preferred in favor of the $\varepsilon$-expansion. In fact, the minimum number of required terms in a divergent series to give good results depends on the perturbation parameter as well as the coefficients in the given series. Thus one can find that perturbative calculations can give very accurate results without any treatment in case the perturbative parameter is small. For instance, in QED the electron anomalous magnetic moment has been deduced from perturbative calculations up to an astonishing accuracy [20, 21]. Recently, the seventh order $g-$expansion of the renormalization group functions for the $O(N)$-symmetric $\phi^4$ model has been obtained and can be analyzed regarding its deterministic power for large $N$. The point is that one can deduce the corresponding seven-loop $\varepsilon$-expansion for which the coefficients decreases with increasing $N$ which gives a hope to mimic the situation of the series in QED. In this work, we will obtain the corresponding seven-loop $\varepsilon-$expansion as a function of $N$ to test its accuracy versus the large-$N$ expansion for $N \geq 4$. To make it clearer, we will show that for a wide rang of $N$, it is sufficient to consider only two-loop or three-loop orders for the prediction of accurate results for the exponents $\nu$ and $\eta$. For the critical exponent $\omega$, however, the Padé approximants can give accurate results.

Ironed by the success of QED perturbative calculations to give unprecedented accuracy, we analyze the seven-loop $\varepsilon-$expansion for the critical exponents within the $O(N)$-symmetric $\phi^4$ model ( for $N \geq 4$) in three dimensions. Our main idea is that as long as for QED where the perturbative parameter is small the series gives accurate results, so it would be also the case for the $O(N)$-symmetric $\phi^4$ model if the perturbative parameter goes to be smaller for larger $N$. In fact both series are divergent but asymptotic. Now consider a series of the form:

$$f(\alpha) = \sum_{i=0}^{\infty} c_i \alpha^i. \tag{1}$$

This series is said to be asymptotic to a function $f(\alpha)$ as $\alpha \to 0$ if [22]

$$f(\alpha) - \sum_{i=0}^{K} c_i \alpha^i \sim c_M \alpha^M, \tag{2}$$

where $c_M$ is the first non-zero coefficient after $c_K$. If the perturbation parameter $\alpha$ is small, the error decreases for small orders and reaches a minimum. In fact, for a generalized Stieltjes series of the form [22]:

$$f(x) = \int_0^{\infty} \frac{\rho(t)}{1+xt} dt \sim \sum_{i=0}^{\infty} (-1)^i a_i x^i, \quad x \to 0^+ \tag{3}$$

where

$$a_i = \int_0^{\infty} t^n \rho(t) dt, \tag{4}$$

and $\rho(t)$ is non-negative for $t > 0$, one can truncate the series just before the smallest term which offers an acceptable numerical estimate of the function represented by the Stieltjes integral above. Such truncation process has been called the optimal truncation rule [22]. Note that, the error of that truncated series is given by:

$$|error(x)| \leq x^{n+1} \int_0^{\infty} t^{n+1} \rho(t) dt = a_{n+1} x^{N+1} \tag{5}$$

which shows that the error getting smaller as $x$ or $a_M$ ( or both) takes smaller values. Moreover, Padé approximation for Stieltjes series has a well known formula of convergence [22, 23]. In fact, the optimal asymptotic approximation also works for asymptotic series which are not of Stieltjes type (see ex.3 pp.94 in Ref.[22]). Besides, the Padé approximations for an asymptotic series are also known to give reliable results when the perturbation parameter is small [24]. For the model at hand, in three dimensions $\varepsilon$ is fixed to one for all $N$ values. However such error formula depends also on $a_{n+1}$ which depends on $\sigma^{n+1} = (\frac{3}{N+8})^{n+1}$. In fact such dependance is clear in all orders in the seven-loop series and extends to large orders. The point is that the RG series of the $\varepsilon-$expansion within the $O(N)$-symmetric $\phi^4$ model has a large-order behavior of the form[2]:

$$c_i \sim \gamma \left(-\sigma\right)^i i^b i! \left(1 + O(\frac{1}{i})\right), \ i \to \infty, \tag{6}$$

where $b$ is amplitude dependent. For the $\eta$ critical exponent $b = 3 + \frac{N}{2}$ while it is $4 + \frac{N}{2}$ for $\nu^{-1}$ and for the approach to scaling critical exponent $\omega$ it is $5 + \frac{N}{2}$. This means that one can replace $\varepsilon$ by the perturbative parameter $\alpha = \sigma\varepsilon$. Accordingly, if the optimal truncation exists, it should give accurate results for large $N$.

Based on the discussions above, truncating the series at the term just preceding the smallest term will lead to a good numerical estimate (see P. 122 in Ref.[22]). Besides, if the perturbation parameter is small then one can even get better estimate by using Padé approximation (especially for intermediate values of $N$).

The rest of this paper will go as follows. In section II, we stress the the critical exponent $\nu$ using optimal truncation and Padé approximations for $N \geq 4$. For the critical exponent $\eta$, we stress it in section III using both optimal truncation as well as Padé approximations too. In both cases and for $N \geq 4$, accurate results have been obtained using only two or three orders of the series which is truncated according the optimal truncation rule. Since the optimal truncation includes only one term for the critical exponent $\omega$ and the series is of Stieltjes type, the Padé approximations are applied which give accurate results. Summary and conclusions follow in section V.

## II. OPTIMAL TRUNCATION FOR THE SERIES OF THE CRITICAL EXPONENT $\nu$ WITHIN THE $O(N)$-SYMMETRIC $\phi^4$ MODEL

The Lagrangian density of the $O(N)$-symmetric $\phi^4$ model is given by:

$$\mathcal{L} = \frac{1}{2}\left(\partial \Phi\right)^2 + \frac{m^2}{2}\Phi^2 + \frac{\lambda}{4!}\Phi^4. \tag{7}$$

$\Phi$ here is a vector with components $(\phi_1, \phi_2, \phi_3, .........\phi_N)$ satisfying an $O(N)$ symmetry such that $\Phi^4 = (\phi_1^2 + \phi_2^2 + \phi_3^2 + ..........\phi_N^2)^2$. Within the minimal subtraction scheme $(\overline{MS})$, the seven loop renormalization group functions have been obtained as [25–27] :

$$\beta(g) = -\varepsilon g + \frac{g^2}{3}(N+8) - \frac{g^3}{3}(3N+14) + .......... + O(g^8),$$

$$\gamma(g) = \frac{g^2}{36}(N+2) - \frac{g^3}{32}(N+2)(N+8) + .......... + O(g^7),$$

$$\gamma_m(g) = \frac{g}{6}(N+2) - \frac{5g^2}{36}(N+2) + \frac{g^3}{72}(N+2)(5N+37) + .......... + O(g^7). \quad (8)$$

$\beta(g)$ is the RG $\beta$−function describing the coupling flow while $\gamma(g)$ is the anomalous field dimension and $\gamma_m(g)$ represents the anomalous mass dimension. The approach to scaling exponent can be obtained from the relation $\omega = \beta'(g_c)$ where $g_c$ is the critical coupling satisfying the condition $\beta(g_c) = 0$. All the above series are divergent with zero radius of convergence and a large order behavior for the coefficients of the form $c_i \sim \tau(-1)^i i! i^b$[2]. The explicit form of the coefficients in all of the above series shows a direct proportionality to $N$ and this is why the power of prediction of the perturbative RG for large $N$ has been overlooked in literature in favor of the large$-N$ expansion.

One can obtain the corresponding $\varepsilon$-expansion for $\gamma_m$ by solving the equation $\beta(g_c) = 0$ for $g_c$ as a function of $\varepsilon$ and then substitute back into $\gamma_m$ equation. Accordingly, one can obtain the seven-loop expansion of $\nu^{-1}$ as :

$$\nu^{-1} = 2 + \gamma_m(g_c) = 2 - \frac{\varepsilon(N+2)}{(N+8)} - \frac{\varepsilon^2(N+2)}{2(N+8)^3}(13N+44) + .......... + O(\varepsilon^7). \quad (9)$$

For $N = 4$, we have the result:

$$\nu^{-1} = 2.00000 - 0.500000\varepsilon - 0.166667\varepsilon^2 + 0.105856\varepsilon^3 - 0.278661\varepsilon^4$$
$$+ 0.702167\varepsilon^5 - 2.23369\varepsilon^6 + 7.97005\varepsilon^7 + O(\varepsilon^8).$$

One can realize that the smallest coefficient exists at the 3-loop order which means that the two-loop partial sum represents the optimal truncation as studied in the introduction section. So, the optimal approximation is given by:

$$\nu^{-1} \simeq 2.00000 - 0.500000\varepsilon - 0.166667\varepsilon^2,$$

where in three dimensions ($\varepsilon = 1$) this gives the result $\nu = 0.75$ which is very competitive to the recent MonteCarlo calculation of $\nu = 0.74817(20)$ in Ref.[3] and the conformal bootstrap (CB) calculations of $\nu = 0.7508(34)$ [28].

For $N = 5$, we generated the seven-loop result as:

$$\nu^{-1} = 2.00000 - 0.538462\varepsilon - 0.173646\varepsilon^2 + 0.0964117\varepsilon^3 - 0.262298\varepsilon^4$$
$$+ 0.622526\varepsilon^5 - 1.89386\varepsilon^6 + 6.42964\varepsilon^7 + O\left(\varepsilon^8\right). \quad (10)$$

Again the best partial sum is given by truncating at the two-loop order or:

$$\nu^{-1} \sim 2.00000 - 0.538462\varepsilon - 0.173646\varepsilon^2,$$

which gives the result $\nu = 0.77646$ compared to the recent MC result $0.7802(6)$ and the non-perturbative renormalization group (NPRG) result of $0.7797(9)$ [29]. The two-loop truncation keeps its superiority up to $N = 65$ but for $N = 66$, truncation at the third order represents the best partial sum. However, since $\sigma = \frac{3}{N+8}$, which exists in all the orders, it is expected that all the seven orders give good results for large $N$. Also, the smallness of $\sigma$ for large-N makes the Padé approximation reliable. For instance, the seven-loop Padé[4/3] approximant yields the result (for $N = 5$) $0.77634$.

For the case $N = 10$, we have

$$\nu^{-1} = 2.00000 - 0.666667\varepsilon - 0.179012\varepsilon^2 + 0.0597939\varepsilon^3 - 0.185638\varepsilon^4$$
$$+ 0.3569416\varepsilon^5 - 0.901457\varepsilon^6 + 2.52130\varepsilon^7 + O\left(\varepsilon^8\right). \quad (11)$$

Again, the two-loop result gives $\nu = 0.86631$ compared the MC result $0.8797(9)$ [3] and the NPRG result $0.8776(10)$ [29]. For this case also the Padé[4/3] approximant for the seven loops gives $\nu = 0.88213$.

In literature, the large-N expansion is thought to produce results that are more accurate than the $\varepsilon$-expansion. The second order for $\nu$ has been obtained to be of the form [2, 8, 9]:

$$\nu = 1 - \frac{4 \times 8}{3\pi^2}\left(\frac{1}{N}\right) + \left(\frac{56}{3} - \frac{9}{2}\pi^2\right)\left(\frac{8}{3\pi^2}\frac{1}{N}\right)^2 + O(N^{-3}), \quad (12)$$

For $N = 4$, this form gives the result $\nu = 0.61234$ while the Padé[1/1] approximant gives $\nu = 0.521975$. It is very clear that the best partial sum of the $\varepsilon$-expansion is more accurate than the second order of the large-$N$ expansion and its Padé[1,1] approximation. Also, a more recent result for the large-$N$ expansion that takes the $2PI$ effective action into account gives [30]:

$$\nu = \frac{1}{1 + \eta + \frac{1}{\pi^2 N A(\eta)}}, \quad (13)$$

where

$$A(\eta) = \frac{1}{8\pi^{\frac{3}{2}}} \frac{\Gamma\left(\frac{1}{2} - \eta\right) \left(\Gamma\left(\frac{1+\eta}{2}\right)\right)^2}{\left(\Gamma\left(1 - \frac{\eta}{2}\right)\right)^2 \Gamma(1 + \eta)},$$

with the critical exponent $\eta$ is satisfying the equation:

$$4\eta(1 - 2\eta \cos(\pi\eta)) = N(2 - \eta)(3 - \eta) \sin^2\left(\frac{\pi\eta}{2}\right). \qquad (14)$$

For $N = 4$, these formulas gives $\nu = 0.79162$. Although the result for $\nu$ seems good but it depends on an inaccurate prediction for $\eta$ ($\eta = 0.0626683$).

According to the above discussion, the two-loop $\varepsilon$-expansion gives accurate results for a wide range of $N \geqslant 4$. However, for large-N, both the optimal truncation of the two-loop $\varepsilon$-expansion expansion as well as the large-N expansion in Eq.(12) gives comparable results as shown in Fig.1.

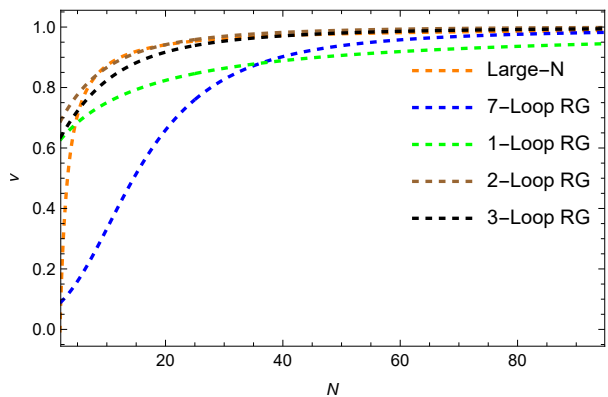

FIG. 1. *In this graph, we plot the $\varepsilon$ series for the critical exponent $\nu$ for different loops. The $1/N$ expansion from Eq.(12) is also plotted for comparison. It is clear that different orders (including those not shown in the figure) and the $1/N$-expansion are all meeting for large $N$.*

TABLE I. The optimal truncation of the $\nu$-series for different $N$ values and the seven-loop Padé[4/3] approximations compared to the predictions from the recent MC [3], CB[28], NPRG [29], Strong-coupling (SC)[2, 31], Padé-Borel resummation (PB) [2, 32] calculations as well as predictions from the large-N expansion in Eq.(12).

| N | Optimal Truncation This work | Padé[4/3] This work | $1/N$-Expansion Eq.(12) | Other Algorithms |
|---|---|---|---|---|
| 4 | 0.75000 | 0.74400 | 0.61234 | $0.74817(20)^{[3]}$ $0.7508(34)^{[28]}$ |
| 5 | 0.77646 | 0.77634 | 0.70867 | $0.7802(6)^{[3]}$ $0.7797(9)^{[29]}$ |
| 10 | 0.86631 | 0.88213 | 0.87313 | $0.8797(9)^{[3]}$ $0.8776(10)^{[29]}$ |
| 20 | 0.941660 | 0.95001 | 0.94126 | $0.938^{[2, 31]}$ $0.930^{[2, 32]}$ |
| 32 | 0.97294 | 0.97818 | 0.96439 | $0.964^{[2, 31]}$ $0.958^{[2, 32]}$ |

In going to larger $N$, the results from the truncated (two-loop) series are more precise. For instance taking $N = 20$, we get the result $\nu = 0.941660$ while the Padé[4/3] approximant results in the value 0.95001. These results can be compared to the strong-coupling results of 0.938 [2, 31] and the Padé-Borel resummation result of $\nu = 0.930$ [2, 32]. For $N = 32$, the best truncation gives the result 0.97294 and the seven-loop Padé[4/3] approximant yields the result 0.97818. The strong-coupling result from Refs.[2, 31] gives the value 0.964 while Padé-Borel resummation in Refs.[2, 32] gives the result 0.958. In table-I, we listed the results of the exponent $\nu$ for different $N$ and compared them to the predictions from other (more sophisticated) techniques.

The cases $6, 8, 10, 12$ have been subjected to recent $MC$ calculation for the quantity $y_t = \nu^{-1}$. So for more comparisons, we consider the case $N = 6$, where the seven-loop result

4is obtained as:

$$\nu^{-1} = 2.00000 - 0.571429\varepsilon - 0.177843\varepsilon^2 + 0.0874661\varepsilon^3 - 0.245698\varepsilon^4$$
$$+ 0.552649\varepsilon^5 - 1.61319\varepsilon^6 + 5.23321\varepsilon^7 + O\left(\varepsilon^8\right). \tag{15}$$

The smallest coefficient is the third order which means that again the two-loop is the best partial sum which results in the result $\nu^{-1} = 1.2507$ compared to the MC result of $\nu^{-1} = 1.2375(9)$( TABLE V in Ref.[3]). The seven-loop Padé[4/3] approximant also gives the result $\nu^{-1} = 1.24265$.

For the case $N = 8$, we have the result:

$$\nu^{-1} = 2.00000 - 0.625000\varepsilon - 0.180664\varepsilon^2 + 0.0719956\varepsilon^3 - 0.213974\varepsilon^4$$
$$+ 0.440222\varepsilon^5 - 1.19138\varepsilon^6 + 3.567301\varepsilon^7 + O\left(\varepsilon^8\right). \tag{16}$$

The two-loop optimal approximation gives $\nu^{-1} = 1.1943$ compared to the MC result is $1.1752(10)$ while the Padé approximant gives the value $\nu^{-1} = 1.17656$.

For the case $N = 12$, we can get the result:

$$\nu^{-1} = 2.00000 - 0.700000\varepsilon - 0.175000\varepsilon^2 + 0.0502971\varepsilon^3 - 0.161056\varepsilon^4$$
$$+ 0.294570\varepsilon^5 - 0.697508\varepsilon^6 + 1.83965\varepsilon^7 + O\left(\varepsilon^8\right), \tag{17}$$

with the two-loop optimal truncation gives the value $\nu^{-1} = 1.125$ while the recent MC result gives $1.1108(17)$[3]) and the Padé[4/3] approximant predicts the value $\nu^{-1} = 1.10506$. In table-II, we listed our optimal approximations (two-loop) as well as the seven-loop Padé[4/3] approximations and compared them with the recent MC calculations and the $1/N$-expansion results for those cases.

TABLE II. The optimal truncation results for $\nu^{-1}$ for $N = 6, 8, 10, 12$ and the seven-loop Padé[4/3] approximations compared to the predictions from the recent MC (2022) calculations and predictions from large-N expansion in Eq.(12).

| N | $\nu^{-1}$ | | | |
|---|---|---|---|---|
| | Optimal Approximation This work | Padé[4/3] This work | 1/N-Expansion Eq.(12) | MC2022 Ref.[3] |
| 6 | 1.2507 | 1.24265 | 1.3027 | 1.2375(9) |
| 8 | 1.1943 | 1.1766 | 1.1968 | 1.1752(10) |
| 10 | 1.1543 | 1.1336 | 1.1453 | 1.1368(12) |
| 12 | 1.1250 | 1.1051 | 1.1150 | 1.1108(17) |

## III. OPTIMAL TRUNCATION FOR THE $\varepsilon$-EXPANSION OF THE $\eta-$EXPONENT

In view of the recent 7-loop g-expansion for the field anomalous dimension $\gamma$, we generated the seven-loop $\eta$-exponent as:

$$\eta = \frac{\varepsilon^2 (N+2)}{2(N+8)^2} + \frac{\varepsilon^3 (N+2)}{8(N+8)^4} \left(-N^2 + 56N + 272\right) + .......... + O\left(\varepsilon^7\right). \qquad (18)$$

For a wide range of $N$, the lowest term is the four-loop contribution which means that the best partial sum is the one truncated at the third order.

For $N = 4$, it gives the result $\eta = 0.0381944$ [3]) compared to MC result of $0.03624(8)$ and the CB result $0.0378(32)$[33] while the NPRG result is $0.0360(12)$[29]. Also, the seven-loop Padé[3/2] approximation yields the result $0.0361286$.

For the case $N = 5$, the three-loop truncation predicts the value $\eta = 0.0368553$. This result is considered to be good in view of the recent MC result of $0.03397(9)$ [3]) and NPRG of $0.03397(9)$ [29] as well as our seven-loop Padé[3/2] approximation which gives the result $0.0346017$.

For $N = 10$, the optimal truncation gives the result $\eta = 0.028978$ and the seven-loop Padé[3/2] approximation gives the value $\eta = 0.0242984$. In Ref.[3], MC calculations gives the result $\eta = 0.02302(12)$ while NPRG result is $\eta = 0.0231(6)$ [29].

In increasing $N$, the parameter $\sigma = \frac{3}{N+8}$ decreases and thus one expects more precise results. For instance when considering the case $N = 20$, the best partial sum is the fourth

order one of the form:

$$\eta \simeq 0.014031\varepsilon^2 + 0.004438\ \varepsilon^3 - 0.004797\ \varepsilon^4, \tag{19}$$

which yields the result 0.0136715 while the seven loop Padé[3/2] approximant gives the value 0.0125074. These predictions can be compared to the strong-coupling algorithm which predicts the value 0.0125[2, 31] and the result 0.014 from Padé-Borel resummation in Ref.[2, 32].

For the case $N = 32$, the four-loop truncation gives $\eta \simeq 0.00862417$ while Padé[3/2] approximant yields the result $\eta = 0.00786012$. The strong-coupling prediction gives the result $\eta = 0.00814$ [2, 31] and Padé-Borel resummation gives the result $\eta \simeq 0.009$ [2, 32]. The above results are listed in table-III where it is also compared to predictions from other techniques.

Needles to say that (order-by order) the large-$N$ limit of $\eta$ is zero which is exact as deduced from the large-$N$ expansion of the form [2, 8, 9]:

$$\eta = \left(\frac{8}{3\pi^2}\frac{1}{N}\right) - \frac{28}{3}\left(\frac{8}{3\pi^2}\frac{1}{N}\right)^2 - \left(\frac{653}{18} - \left(27\log 2 + \frac{47}{4}\right)\zeta(2) + \frac{189}{4}\zeta(3)\right)\left(\frac{8}{3\pi^2}\frac{1}{N}\right)^3 + O(N^4) \tag{20}$$

TABLE III. The optimal truncation results for $\eta$ with different $N$ values and the seven-loop Padé[3/2] approximations compared to the predictions from the recent MC [3], CB[28], NPRG [29], Strong-coupling (SC)[2, 31], Padé-Borel resummation (PB) [2, 32] calculations and predictions from large-N expansion in Eq.(20). According to the optimal truncation rule, for $N = 4, 5, 10$, we truncated the series at the third order while for $N = 20, 32$ the truncation exists at the fourth loop order.

| N | $\eta$ | | | |
|---|---|---|---|---|
|   | Optimal Approximation<br>This work | Padé[3/2]<br>This work | $1/N$-Expansion<br>Eq.(20) | Different Algorithms |
| 4 | 0.0381944 | 0.0361286 | 0.011722 | 0.03624(8)[3]<br>0.0378(32)[33] |
| 5 | 0.0368553 | 0.0346017 | 0.020004 | 0.03397(9)[3]<br>0.0338(11)[29] |
| 10 | 0.0289781 | 0.0242984 | 0.019358 | 0.02302(12)[3]<br>0.0231(6)[29] |
| 20 | 0.0136715 | 0.0125074 | 0.0117 | 0.0125[2, 31]<br>0.014[2, 32] |
| 32 | 0.00862417 | 0.00786012 | 0.0077522 | 0.00814[2, 31]<br>0.009[2, 32] |

The dependance on $N$ can be realized from Fig.2 where we plotted $\eta$ versus $N$ for different orders of $\varepsilon$-expansion and the large-$N$ expansion in Eq.(20). Again, like the case of $\nu$ exponent, different orders as well as the large-$N$ expansion coincide at large $N$.

Now we consider the cases $N = 6, 8, 12$ to compare with the recent MC calculations ( second column in TABLE V of Ref.[3]). For $N = 6$, the three-loop result is 0.0352978 compared to $\eta = 0.03157(14)$ from MC calculations [3] while our Padé[3/2] gives the result 0.0327137. For the case $N = 8$, we get the three-loop result of 0.0320435 while the MC result is 0.02675(15) and the Padé[3/2] approximant yields the result 0.0284581. In table-IV we list our optimal truncation approximation and seven-loop Padé[3/2] approximations results in comparison with these recent MC calculations and large-$N$ expansion.

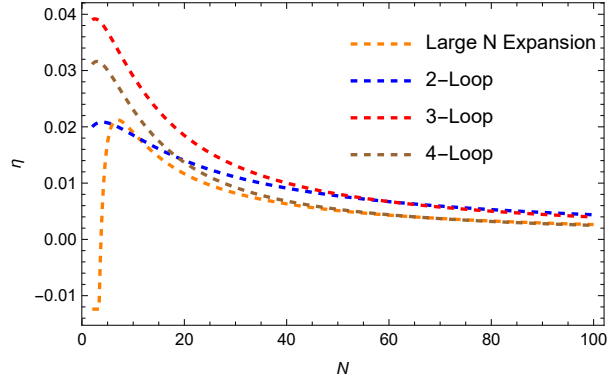

FIG. 2. *Here we plot the critical exponent $\eta$ for different orders of the RG series. The $1/N$ expansion from Eq.(20) is also plotted for comparison. It is clear that different orders (including those not shown in the figure) and the $1/N$-expansion are all meeting for large $N$.*

TABLE IV. The optimal truncation results for $\eta$ for $N = 6, 8, 10, 12$ values and the seven-loop Padé[3/2] approximations compared to the predictions from the recent MC (2022) calculations and predictions from large-N expansion in Eq.920).

| | $\eta$ | | | |
|---|---|---|---|---|
| N | Optimal Approximation This work | Padé[3/2] This work | 1/N-Expansion Eq.(12) | MC 2022 Ref.[3] |
| 6 | 0.0352978 | 0.0327137 | 0.022182 | 0.03157(14) |
| 8 | 0.0320435 | 0.0284581 | 0.021472 | 0.02675(15) |
| 10 | 0.0289781 | 0.0242984 | 0.019358 | 0.02302(12) |
| 12 | 0.0205414 | 0.0207401 | 0.017294 | 0.0199(3) |

## IV. SEVEN-LOOP $\varepsilon$−EXPANSION FOR THE CRITICAL EXPONENTS $\nu, \eta$ AND $\omega$ OF THE $O(N)$-SYMMETRIC $\phi^4$ MODEL

In the previous sections, we showed that the best truncated series produces accurate results for the the exponents $\nu, \eta$. The results are more precise for larger $N$ and are going to be exact at the large-$N$ limit. For the critical exponent $\omega$, however, the story is different. To analyze it, we first obtain the seven-loop order as:

$$\omega = \varepsilon - \frac{3\varepsilon^2}{(N+8)^2}(3N + 14) + \ldots\ldots + O\left(\varepsilon^8\right). \qquad (21)$$

This series (order by order) predicts exact result as $N \to \infty$ (see Fig.3). However, for a wide range of $N$, optimal truncation includes only one-loop contribution and thus results in inaccurate results. Starting at $N = 140$, we have a best truncation at the two-loop order or:

$$\omega \simeq 1 - \frac{3}{(N+8)^2}(3N+14).$$

This optimal truncation gives for $N = 140$ the result $\omega \simeq 0.940\,56$. For the Large-$N$

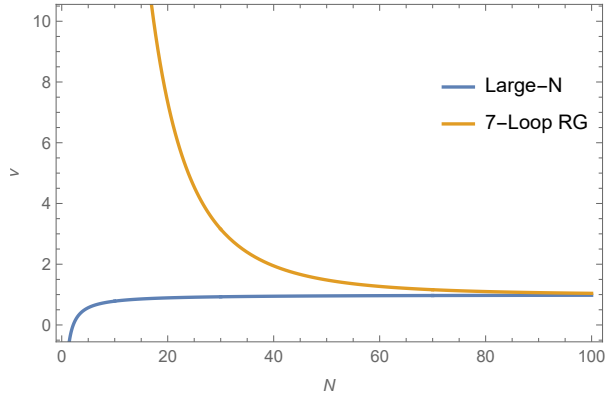

FIG. 3. *In this graph, we plot the 7-loop $\varepsilon$-series for the critical exponent $\omega$ and compare it with the $1/N$ expansion from Eq.(21). It is clear that both are close to each other only for large $N$.*

expansion, the result is in the from [2, 8, 9]:

$$\omega = 1 - \frac{8 \times 8}{3\pi^2}\left(\frac{1}{N}\right) + 2\left(\frac{104}{3} - \frac{9}{2}\pi^2\right)\left(\frac{8}{3\pi^2}\frac{1}{N}\right)^2 + O(N^{-3}), \qquad (22)$$

where for $N = 140$ it gives the result for $\omega \simeq 0.98449$. On the other hand the seven-loop Padé[3/3] gives the result $0.985543$. Accordingly, for a wide range of $N$, the RG $\varepsilon$-expansion is in a need for a treatment in order to be able get reliable results especially for $N < 140$. The RG $\varepsilon$-expansion for the exponents $\nu, \eta$ and $\omega$ can be written as

$$f(\alpha) = \sum_{i=0}^{\infty} c_i \alpha^i,$$

where $\alpha = \sigma\varepsilon = \frac{3\varepsilon}{N+8}$. One can realize that in all the seven orders at hand, the factor $\sigma^i$ is appearing which means that the perturbation parameter $\alpha$ is taking smaller values for larger $N$ and thus Padé[3/3] approximants for the seven-loop series can predict accurate results [24] that are very competitive to rigorous techniques like MC, CB and NPRG. For the $\omega/\varepsilon$ series, however, there exists alternation on the sign of different terms. Let us consider a specific example for $N = 4$ where the series takes the form:

$$\frac{\omega}{\varepsilon} = 1 - 0.541667\varepsilon + 1.15259\varepsilon^2 - 3.27193\varepsilon^3 + 10.8016\varepsilon^4 - 40.5665\varepsilon^5 + 166.256\varepsilon^6. \quad (23)$$

Two main realizations about this series. According to the optimal truncation rule, the best truncated series is $\omega \approx 1$ which is not acceptable at all specially for relatively small $N$ values. The second important realization is that the series alternates in sign which means it is a series of Stieltjes [34]. Such type of series has a well-known convergence formula for the Padé approximants [23, 24]. To value the Padé calculations for $N \geq 4$, let us compare it with the recent Borel with conformal mapping calculations for the six-loop series [5]. For $N = 4$, the result for the six-loop $\varepsilon$-expansion is $\nu = 0.7397(35)$ while our six-loop Padé result is $\nu = 0.740814$. For $\eta$, the six-loop result in Ref.[5] is $0.0366(4)$ while our Padé result is $0.0359019$ for the same order. Similarly, for the critical exponent $\omega$, our six-loop Padé result is $0.782122$ while for the same order the Borel (with conformal mapping) result is $0.794(9)$. These results are showing how seriously the Padé method is to be taken when treating RG series for $N \geq 4$. In view of these analysis, we generated the seven-loop Padé results for the three exponents which are listed in table-V and compared them with different predictions from other tools.

TABLE V. The seven-loop Padé approximations results for the critical exponets $\nu, \eta, \omega$ for $N = 4, 5, 10$ compared to the predictions from the recent MC [3] calculations, CB[28, 33], NPRP [29], Strong-coupling (SC)[2, 31] and Padé-Borel resummation (PB) [2, 32].

| N | $\nu$ | $\eta$ | $\omega$ |
|---|---|---|---|
| 4 | 0.74400[This work] | 0.0361286[This work] | 0.782882[This work] |
|   | 0.74817(20)[3] | 0.03624(8)[3] | 0.755(5)[3] |
|   | 0.7508(34)[28] | 0.0378(32)[33] | 0.817(30)[28] |
|   | 0.737[2, 31] | 0.031[2, 31] | 0.795[2, 31] |
|   | 0.738[2, 32] | 0.036[2, 32] | 0.788[2, 32] |
| 5 | 0.77634[This work] | 0.0346017[This work] | 0.775261[This work] |
|   | 0.7802(6)[3] | 0.03397(9)[3] | 0.7802(6)[3] |
|   | 0.7797(9)[29] | 0.0338(11)[29] | 0.754(7)[29] |
|   | 0.767[2, 31] | 0.0295[2, 31] | 0.938[2, 31] |
|   | 0.766[2, 32] | 0.034[2, 32] | 0.930[2, 32] |
| 10 | 0.88213[This work] | 0.0242984[This work] | 0.792877[This work] |
|   | 0.8797(9)[3] | 0.02302(12)[3] | 0.816(16)[3] |
|   | 0.8776(10)[29] | 0.0231(6)[29] | 0.807(7)[29] |
|   | 0.866[2, 31] | 0.0216[2, 31] | 0.824[2, 31] |
|   | 0.859[2, 32] | 0.0.024[2, 32] | 0.822[2, 32] |

## V.  SUMMARY AND CONCLUSIONS

The series ($\varepsilon$-expansion) for the critical exponents with in the $(O(N))$-symmetric field theory takes the from $f(\alpha) = \sum_{i=0}^{\infty} c_i \alpha^i$ where $\alpha = \sigma\varepsilon$. Since $\sigma = 3/(N+8)$, the perturbation parameter $\alpha$ gets smaller for larger $N$ values. The series are divergent but asymptotic like the ones in QED. However, in QED, because of the smallness of the fine structure constant one is able extract precise results from partial sum of the given series. This fact is reflected in the electron anomalous magnetic moment which has been extracted from perturbative calculations very accurately [20, 21]. For the $O(N)$-symmetric field theory as $\alpha$ is getting smaller for larger $N$, then one might expect the existence of accurate partial sums for the $\varepsilon$-series.

We obtained the seven-loop $\varepsilon$-series as a function of $N$ and found that term by term the values take their exact form at the large-$N$ limit. This result is celebrated by the non-perturbative $1/N-$Expansion [4, 8–12]. However, the optimal truncation of the $\varepsilon$-series gives more accurate results especially for relatively small $N$ values.

If the perturbation parameter is small, an optimal truncation at the term just preceding the lowest term exists [22] where the truncated series can give good approximation for the asymptotic series. We applied the optimal truncation rule for the three exponents $\nu, \eta$ and $\omega$ for $N \geq 4$. The truncation for the $\nu$ exponent exists at the two-loop order for a wide range of $N$ and for relatively large $N$ the truncation starts to exist at the third order. Such truncation results in competitive results when compared to recent more sophisticated techniques like the MC,CB, NPRG calculations [3, 28, 29]. Likewise, for the critical exponent $\eta$, the optimal truncation exists however at the third and then fourth order which again resulted in accurate results for a wide range of $N$. For the critical exponent $\omega$, however, the optimal truncation exists at the first order which results for a fixed value $\omega = 1$ for a wide range of $N$ values. accordingly, such series needs to be followed by a treatment technique like Padé approximants (for instance).

For an asymptotic series with small perturbation parameter, Padé approximants are thought to give accurate results [24]. In fact, specialty for the $\omega$-series, it is a series of Stieltjes which is well-known to have a convergence formula for the Padé approximants. We applied the Padé approximation for the three series (seven-loop) and found that it gives accurate results. One can take an idea about the accuracy of the Padé approximation for $N = 4$ by comparing its six-loop prediction with the recent Borel with conformal mapping calculations in Ref.[5]. For instance, our Padé[3/3] approximation for $\nu$ gives the result 0.740814 compared to the Borel six-loop result of 0.7397(35) [5]. In fact in comparing to recent MC and CB results, our six-loop Padé[3/3] is more accurate. Also, our six-loop results for $\eta$ is 0.0359019 compared to 0.0366(4) from [5]. This result is very acceptable as we did not use any auxiliary parameters that one has to optimize in order to minimize the error. For $\omega$, we get the result 0.782122 compared to the six-loop resummation prediction of $\omega = 0.794(9)$ from Ref.[5].